\documentclass[vecphys]{svmult}

\usepackage{makeidx}         % allows index generation
\usepackage{graphicx} 
\usepackage{amsmath}       % standard LaTeX graphics tool
\usepackage{multicol}        % used for the two-column index
\usepackage{cite}            % adjusts the "syntax" of the refs in the
\usepackage[bottom]{footmisc}% places footnotes at page bottom
\usepackage{color}  %KWM added this for keywords
\makeindex             % used for the subject index
\newcommand{\nbar}[0]{\bar{n}}
\newcommand{\idx}{\textcolor{black}}
\begin{document}

\title{Weak Measurement and Feedback in Superconducting Quantum Circuits}
\author{K. W. Murch \and R. Vijay \and I. Siddiqi}
\institute{Department of Physics, Washington University, St. Louis, MO, USA
\texttt{murch@physics.wustl.edu}
\and Tata Institute of Fundamental Research, Mumbai, India \texttt{rvijay26@gmail.com}
\and Quantum Nanoelectronics Laboratory, Department of Physics, University of California, Berkeley, CA, USA \texttt{irfan@berkeley.edu}}

% Use the package "url.sty" to avoid problems with special characters
% used in your e-mail or web address.K
% Addresses should be removed from contribution and entered into
% blist.tex" (by the compiler).

\titlerunning{Quantum Trajectories}
\maketitle

\begin{abstract}

We describe the implementation of weak \idx{quantum measurements} in superconducting qubits, focusing specifically on transmon type devices in the circuit quantum electrodynamics architecture.  To access this regime, the readout cavity is probed with on average a single microwave photon. Such low-level signals are detected using near quantum-noise-limited superconducting parametric amplifiers. \idx{Weak measurements} yield partial information about the quantum state, and correspondingly do not completely project the qubit into an eigenstate. As such, we use the measurement record to either sequentially reconstruct the quantum state at a given time, yielding a \idx{quantum trajectory}, or to close a direct \idx{quantum feedback} loop, stabilizing Rabi oscillations indefinitely.  

\end{abstract}

\noindent

Measurement-based feedback routines are commonplace in modern electronics, including the thermostats regulating the temperature in our homes to sophisticated motion stabilization hardware needed for the autonomous operation of aircraft. The basic elements present in such classical feedback loops include a sensor element which provides information to a controller, which in turn steers the system of interest toward a desired target. In this paradigm, the act of sensing itself does not \textit{a priori} perturb the system in a significant way. Moreover, extracting more or less information during the measurement process does not factor into the control algorithm.  For quantum coherent circuits, these basic assumptions do not hold. The act of measurement is invasive, and the so-called backaction drives a system toward an eigenstate of the measurement operator. Furthermore, the information content of the measurement determines the degree of backaction imparted. For example, strong measurements completely project a superposition state into a define eigenstate while extracting enough information to unambiguously allow an observer to determine which eigenstate has been populated. Weak measurements on the other hand, accrue less information, indicating which eigenstate is more likely to be populated, and have proportionally weaker backaction that does not completely collapse a quantum superposition. This type of measurement provides a natural route to implement active feedback in quantum systems with the weak measurement outcome providing the input for a quantum controller that takes into account the measurement induced backaction.

In this chapter, we first discuss a general formalism based on positive operator-valued measures (POVMs) which can describe both weak and strong measurements. We then apply this formalism to a superconducting two-level system coupled to a microwave frequency cavity, which is well approximated by the Jaynes-Cummings model commonly used in cavity quantum electrodynamics. In such a system, the measurement strength for a given integration time can be adjusted by varying the number of photons in the cavity. To access the weak measurement regime, the cavity is typically populated, on average, with less than one photon, requiring an ultra-low-noise amplifier for efficient detection. In contemporary experiments, this function is realized using superconducting parametric amplifiers. After briefly describing these devices, we discuss two basic types of experiments. In the first set, the result of a sequence of weak measurements is used to reconstruct individual quantum trajectories using a Bayesian update procedure. The statistical distribution of an ensemble of many such trajectories is then analyzed to experimentally and theoretically obtain the most likely path. In these experiments, the backaction results either solely from the measurement process or from the combination of measurement and unitary evolution under a coherent drive. In principle, the reconstructed state and the detector values corresponding to the most likely path can be used for arbitrary feedback protocols and optimal control. In the second type of experiment, we use the weak measurement outcome to directly complete an analog feedback circuit. In particular, we demonstrate the real-time stabilization of Rabi oscillations resulting in the suppression of their ensemble decay. Finally, we close with a discussion of future directions for hardware improvement and additional experimentation, particularly in multi-qubit systems.     
 
\section{Generalized Measurements}
\label{sec:1}

In quantum mechanics, predictions about the outcome of experiments are given by \idx{Born's rule} which for a state vector $|\psi_i\rangle$ provides the probability $P(a)=|\langle a|\psi_i\rangle|^2$ that a measurement of an observable described by an operator $\hat{A}$ with eigenstates $|a\rangle$ yields one of the eigenvalues $a$. As a consequence of the measurement, the quantum state is projected into the state $|a\rangle$. Here we consider a qubit, with states $|0\rangle$ and $|1\rangle$, which is conveniently described by the Pauli matrix algebra with pseudo-spin operators $\sigma_x$, $\sigma_y$, and $\sigma_z$.  For example, if we prepare the qubit in an initial state, \begin{eqnarray}
|\psi_i\rangle = |+x\rangle \equiv \frac{1}{\sqrt{2}} (|0\rangle + |1\rangle) \label{eq:x}
\end{eqnarray}
Born's rule tells us the probability of finding the qubit in the state $|0\rangle$, $P(+z) = 1/2$.  Here we have introduced the notation $\pm x$, $\pm z$ to indicate the eigenstate of the $\sigma_x$ and $\sigma_z$ pseudo-spin operators. After the measurement, the qubit will remain in the eigenstate corresponding to the eigenvalue and subsequent measurements will yield the same result.  In this way the measurement changes $\langle \sigma_z \rangle$ from 0 to 1, while changing $\langle \sigma_x \rangle$ from 1 to 0.  This change, associated with a projective measurement of $\sigma_z$ is the backaction of measurement and also referred to as the collapse of the wavefunction.  

The fact that the measurement takes $\langle \sigma_x \rangle$ from 1 to 0 is a consequence of the \idx{Heisenberg uncertainty} principle. Because $\sigma_x$ and $\sigma_z$ do not commute, if one component of the pseudo-spin is known then the others must be maximally uncertain.  Thus some amount of backaction occurs in any measurement, and measurements (such as the projective measurement discussed above) that cause no more backaction than the amount mandated by the Heisenberg uncertainty principle are said to be quantum non-demolition.

In this chapter, we are concerned with a more general type of measurement that can be performed on the system \cite{wisebook,jaco06,brun02,nielsen,silb05,smit06}. Real measurements take place over a finite amount of time and we are interested in describing the evolution of the system along the way.  As such, a projective measurement is composed of a sequence of \idx{partial measurements}. Formally, we can describe these measurements by the theory of \idx{positive operator-valued measures} \cite{wisebook,jaco06,brun02,nielsen}, (POVMs) which yields the probability $P(m)=\mathrm{Tr}(\Omega_m \rho \Omega_m^\dagger)$ for outcome $m$, and the associated \idx{backaction} on the quantum state, $\rho \rightarrow \Omega_m \rho \Omega_m^\dagger/P(m)$ for a system described by a density matrix $\rho$.  POVMs are ``positive" simply because they describe outcomes with positive probabilities and ``operator-valued'' because they are expressed as operators.   These operators $\Omega_m$ obey  $\sum_m \Omega^\dagger_m \Omega_m = \hat{I}$ as is necessary for POVMs.  When $\Omega_a = |a\rangle \langle a|$ is a projection operator and $\rho = |\psi\rangle\langle \psi|$, the theory of POVMs reproduces Born's rule.  For the case described above, the projective measurement of $\sigma_z$ is described by the POVM operators $\Omega_{+z} =(\hat{I}+\sigma_z)/2 $ and $\Omega_{-z} = (\hat{I} - \sigma_z)/2$ and straightforward application of the theory reproduces the effects of projective measurement discussed above.

\subsection{Indirect Measurements}
\label{sec:1:1}

While it is tempting to connect a quantum system directly to a classical measuring apparatus, such an arrangement often results in far more backaction on the quantum system than is dictated by the Heisenberg uncertainty principle. For example, an avalanche photodiode can be used to detect the presence of a single photon, but it achieves this detection by absorbing the photon, thereby completely destroying the quantum state.  In practice, classical devices are composed of too many noisy degrees of freedom to perform QND measurements. 

To circumvent this problem, we break the measurement apparatus into three parts to conduct an \idx{indirect measurement}. The three parts of the apparatus are the quantum system of interest, a quantum \idx{pointer} system that can be coupled to the quantum system, and then a classical measurement apparatus that can be used to record the pointer system.  To execute a measurement, the pointer system is coupled to the quantum system and after sufficient interaction the two systems become entangled.  Then, a classical measurement apparatus, such as a photodetector or an ammeter, is used to make measurements on the pointer system.  These measurements on the pointer system are projective and depending on their outcome, the quantum system's state is accordingly changed.

To illustrate this process consider a system consisting of two qubits \cite{brun02}. We initialize one qubit (the system qubit) in the state in (\ref{eq:x}) and allow a second (environment) qubit to interact with the primary qubit for a brief period of time such that the two qubits are in the entangled state, 
\begin{multline}
|\Psi\rangle \propto \left( (1+\epsilon)|0\rangle_\mathrm{sys} + (1-\epsilon)|1\rangle_\mathrm{sys} \right) \otimes |0\rangle_\mathrm{env}\\
+ \left( (1-\epsilon)|0\rangle_\mathrm{sys} + (1+\epsilon)|1\rangle_\mathrm{sys} \right) \otimes |1\rangle_\mathrm{env}.
\end{multline}
We then make a projective measurement of the second qubit (using the POVM $\Omega_{\pm z}$), and if the result is the $| 0\rangle$ state, then the system qubit is left in the state,
\begin{eqnarray}
|\Psi\rangle \propto  (1+\epsilon)|0\rangle_\mathrm{sys} + (1-\epsilon)|1\rangle_\mathrm{sys}.
\end{eqnarray}
For $\epsilon\ll1$ this partial measurement slightly drives the system qubit to the ground state.  However, what happens if we instead measure the environment qubit in the $\sigma_y$ basis, $| y_\pm \rangle  = \frac{1}{\sqrt{2}} (| 0\rangle \pm i |1\rangle$?  We can express the joint state in this basis, 
\begin{multline}
|\Psi\rangle \propto \left( (1+\epsilon-i(1-\epsilon))|0\rangle_\mathrm{sys} + (1-\epsilon - i(1+\epsilon))|1\rangle_\mathrm{sys} \right) \otimes |y_+\rangle_\mathrm{env}\\
+ \left( (1-\epsilon-i(1+\epsilon))|0\rangle_\mathrm{sys} + (1+\epsilon-i(1-\epsilon))|1\rangle_\mathrm{sys} \right) \otimes |y_-\rangle_\mathrm{env}.
\end{multline}
By factoring out a global phase factors $\phi_+ = \pi/4+\epsilon$ and $\phi_- = \pi/4-\epsilon$ and assuming that $\epsilon\ll1$, the joint state is, 
\begin{multline}
|\Psi\rangle \propto e^{i \phi_+} \left( |0\rangle_\mathrm{sys} + e^{-2 i \epsilon}|1\rangle_\mathrm{sys} \right) \otimes |y_+\rangle_\mathrm{env}\\
+ e^{i\phi_-} \left( |0\rangle_\mathrm{sys} + e^{2i \epsilon}|1\rangle_\mathrm{sys} \right) \otimes |y_-\rangle_\mathrm{env}
\end{multline}
which shows that the system qubit state obtains a slight rotation depending on the detected state of the environment qubit. This simple model indicates two salient features of partial measurements.  First, if the environment and the system are weakly entangled, then projective measurements on the environment cause only small changes in the qubit state, and second, the choice of measurement basis for the environment gives different conditional evolution for the system.  After the projective measurement on the environment qubit the system is in a product state with the environment qubit and no entanglement remains. 
 
We now imagine many such successive interactions between the system qubit and a set of environment qubits.  In this case each entangling interaction and projective measurement of the environment causes a small amount of random backaction on the qubit, which results in a diffusive trajectory for the state of the qubit.  Such weak measurements have been recently implemented using coupled superconducting qubits \cite{groe13}.
 
\subsection{Continuous Measurement}%
\label{sec:1:2}

The measurements that are realized in the cQED architecture described later in section \ref{sec:2} can be represented by a set of POVMs $\{ \Omega_V\}$, where $V$ is the dimensionless measurement result which is scaled so that it takes on average values $\pm 1$ for the qubit in states $\pm z$ respectively,
  \begin{eqnarray}
%\Omega_V = \left(8 k \eta dt/\pi \right)^{1/4} e^{(-4 \eta k dt (V- \sigma_z/2)^2)}
\Omega_V = \left(2 \pi a^2 \right)^{-1/4} e^{(-(V- \sigma_z)^2/4a^2)}.
\label{povm}
\end{eqnarray}
Here,  $1/4a^2 =  k \eta \Delta t$, and $k$ parametrizes the strength of the measurement, $\eta$ is the quantum efficiency, and $\Delta t$ is the duration of the measurement.  The operators $\Omega_V$ satisfy $\int \Omega_V^\dagger \Omega_V dV = \hat{I}$ as expected for POVMs. The probability of each measurement yielding a value $V$ is $P(V) = \mathrm{Tr}(\Omega_V \rho_t \Omega_V^\dagger)$, which is the sum of two Gaussian distributions with variance $a^2$ centered at $+1$ and $-1$ and weighted by the populations $\rho_{00}$ and $\rho_{11}$ of the two qubit states. The $\sigma_z$ term in $\Omega_V$ causes the back action on the qubit degree of freedom, $\rho \rightarrow \Omega_V \rho \Omega_V^\dagger/P(V)$, due to the readout of the measurement result $V$. Eq. (\ref{povm}) can also describe a stronger measurement, ultimately yielding the limit where the two Gaussian distributions are disjoint, and the readout projects the qubit onto one of its $\sigma_z$ eigenstates, with probabilities $\rho_{00}$ and $\rho_{11}$. The strength of the measurement is controlled by the parameter $k$ and we will see later in section \ref{sec:2} how this quantity is related to experimental parameters.

For \idx{weak measurements}, $\Delta t\ll \tau_c $, where  the characteristic measurement time $\tau_c = 1/4 k \eta$ describes the time it takes to separate the Gaussian measurement histograms by two standard deviations. The distribution of measurement results is then approximately given by a single Gaussian that is centered on the expectation value of $\sigma_z$, 
\begin{eqnarray}
P(\Omega_V) \simeq e^{-4 k \eta \Delta t (V-\langle \sigma_z \rangle)^2},
\end{eqnarray}
which highlights the fact that $V$ is simply a noisy estimate of $\langle \sigma_z \rangle$. This allows the measurement signal to be written as the sum of $\langle\sigma_z\rangle$ and a zero mean Gaussian random variable   \cite{jaco06}.   %As such, $V$ can be written as a sum of $\langle \sigma_z \rangle$ and a stochastic quantity \cite{jaco06}, $V = \langle \sigma_z\rangle + \Delta W/(\sqrt{8 k \eta} \Delta t).$ {\bf [Note that the fact that $V$ is proportional to $\langle \sigma_z\rangle$ gives a good intuitive picture for measurement pinning in the bayesian picture]} 

The time evolution of the quantum state following a sequence of measurements described by Eq. (\ref{povm}) in the limit $\Delta t\rightarrow 0$  is given by the \idx{stochastic master equation} \cite{wisebook}:
\begin{eqnarray}
\frac{d \rho}{dt} =  k(\sigma_z \rho \sigma_z - \rho) 
+ 2 \eta k( \sigma_z \rho +\rho \sigma_z - 2 \mathrm{Tr}(\sigma_z \rho) \rho) V(t). \label{eq:rho}
\end{eqnarray}
Here, the first term is the standard master equation in Lindblad form and the second  term is the stochastic term that updates the state based on the measurement result.  In the case of unit detector efficiency ($\eta= 1$) the stochastic master equation (perhaps surprisingly) yields pure state dynamics. This happens because the decoherence term with rate $k$ in the equation is exactly compensated by the stochastic term to yield a random pure state evolution; the first term shrinks the Bloch vector towards the $z$-axis, while the stochastic term adds a random transverse component, putting it back to the surface of the Bloch sphere again. In contrast, if we set $\eta = 0$, all random backaction is suppressed, and we instead obtain a conventional, deterministic master equation, where probing of the system causes extra Lindblad decoherence on the system, but yields no information.

\section{Quantum Measurements in the cQED Architecture}
\label{sec:2}

The circuit QED architecture \cite{cQEDtheory} provides both excellent coherence properties and high fidelity measurement. In particular, it is an ideal test bed for studying quantum measurements because it enables one to implement text book quantum measurements relatively easily. The two key aspects which make this possible is the applicability of an ideal measurement Hamiltonian \cite{cQEDtheory} and the availability of quantum limited parametric amplifiers \cite{cast08, hatr11para}. 

\begin{figure}[t]
\centering
\includegraphics*[width=.7\textwidth]{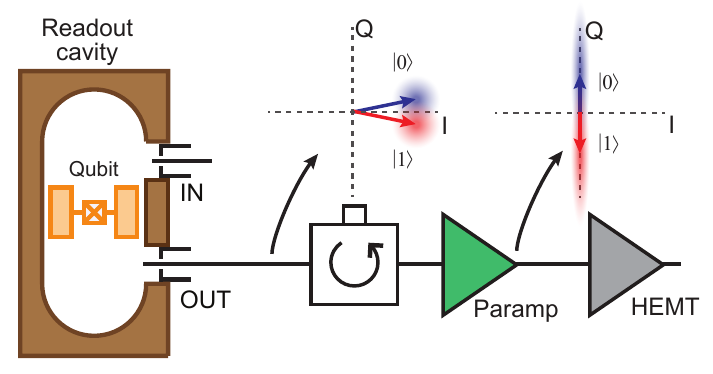}
%\picplace{5cm}{2cm} % Give the correct figure height and width in cm
\caption[]{Circuit QED setup consisting of a qubit dispersivley coupled to a readout cavity.  A signal transmitting the cavity at the cavity frequency aqcuires a qubit-state-dependent phase shift, shown as two phasors in the $I-Q$ plane.  Phase sensitive amplification amplifies one quadrature exclusivley. }
\label{fig:cqed}       % Give a unique label
\end{figure}

\idx{Circuit QED} (cQED) is essentially the implementation of the cavity QED architecture \cite{berm94book} using superconducting circuits. Instead of an atom interacting with the electromagnetic field inside a Fabry-Perot cavity, the cQED architecture uses a superconducting qubit as an `artificial' atom which interacts coherently with the electromagnetic field in an on-chip waveguide resonator \cite{wallraff_nature} or a 3D waveguide cavity \cite{paik113D}. Figure \ref{fig:cqed} shows the basic cQED setup. Dispersive readout in cQED was introduced in Chapter 6 and we briefly reintroduce it here to make this chapter more self contained.   The Hamiltonian describing this interaction is the Jaynes-Cummings Hamiltonian, 

 \begin{eqnarray}
H =\frac{\hbar \omega_{01}}{2}\sigma_z +  \hbar \omega_c (a^\dagger a  + \frac{1}{2})+ \hbar g (a \sigma^+ + a^\dagger \sigma_-)     \label{eq:JCH}
\end{eqnarray}
where the first term represents the qubit as a pseudo-spin, the second term is the cavity mode, and the third term represents the interaction between the qubit and the electromagnetic field in the rotating wave approximation. Here $\omega_{01}$ is the transition frequency between the qubit levels, $\omega_c$ is the cavity mode frequency, $g$ is the coupling strength between the qubit and cavity mode and $\sigma_{\pm} = (\sigma_x \pm i\sigma_y)/2$ are the qubit raising and lowering operators. Typically, the qubit frequency is far detuned from the cavity frequency to protect the qubit from decaying into the mode to which the cavity is strongly coupled. In this dispersive regime ($\Delta = \omega_{01}- \omega_c \gg g$), the effective Hamiltonian reduces to,

 \begin{eqnarray}
H =\frac{\hbar \omega_{01}}{2}\sigma_z +  \hbar (\omega_c  + \chi \sigma_z) (a^\dagger a  + \frac{1}{2})  \label{eq:JCDH}
\end{eqnarray} where $\chi$ is called the \idx{dispersive shift}. It is now possible to see how the measurement is implemented. The cavity mode frequency is a function of the qubit state with $\chi$, which depends on $g$ and $\Delta$, setting the magnitude of this shift. The measurement proceeds by probing the cavity with a microwave signal and detecting the qubit state dependent phase shift of the scattered microwave signal. In this architecture, one can clearly see the process of indirect measurement introduced earlier. The \idx{pointer system} is the microwave field which interacts with the qubit and gets entangled with it. One then measures the output microwave field which in turn determines the qubit state. As mentioned earlier, the exact measurement on the pointer determines the backaction on the qubit and that can be controlled by the choice of amplification method used to detect the output microwave field. We will however come to that a little later.

\subsection{Dispersive Measurements}
\label{sec:2:1}

\begin{figure}[t]
\centering
\includegraphics*[width=.8\textwidth]{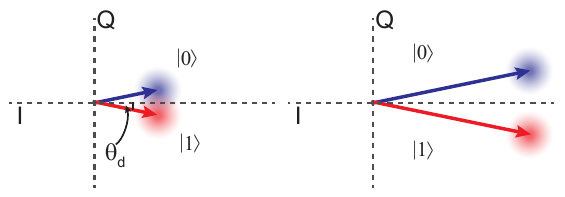}
%\picplace{5cm}{2cm} % Give the correct figure height and width in cm
\caption[]{Output coherent states represented in the IQ plane. The length of the vector increases with increasing photon number in the coherent state and leads to better distinguishability between the qubit states. }
\label{fig:iqplane}       % Give a unique label
\end{figure}

We will now formalize the interaction between the qubit and the pointer state using the language of coherent states to describe the quantum state of the microwave field. Let $|\alpha\rangle$ represent the coherent state of the microwave field sent to the cavity with an average photon number $\bar{n} = |\alpha|^2$. The initial state of the qubit is $a_0|0\rangle + a_1|1\rangle$. After interaction, the final entangled state of the system is given by
\begin{eqnarray}
|\psi_f\rangle = a_0|e^{-i \theta_d}\alpha\rangle|0\rangle + a_1|e^{+i \theta_d}\alpha\rangle|1\rangle \label{eq:qubitpointer}
\end{eqnarray} where $\theta_d$ is the dispersive phase shift of the scattered microwave signal. The two coherent states can be represented in the quadrature plane as shown in figure \ref{fig:iqplane}. For a fixed dispersive phase shift $\theta_d$, the larger the average number of photons in the coherent states, the easier it is to distinguish them. Here, the distinguishability of the output coherent states is directly related to the distinguishability of the qubit state i.e. the strength of the measurement can be controlled by the microwave power.  

One can now choose to measure the output microwave field by performing a homodyne measurement and choosing one of the two possible quadratures. From figure \ref{fig:iqplane}, it is clear that the qubit state information is encoded in the `Q' quadrature while the `I' quadrature has the same value for both qubit states. We will first describe  the single quadrature measurement of the `Q' quadrature using a phase-sensitive amplifier and the associated backaction on the qubit. Since the coherent state is not a two-level state, a measurement on it yields a continuous range of values and corresponds to the case discussed in section \ref{sec:1:2}. Typically, experiments with microwave signals employ mixers operating at room temperature to implement homodyne detection. However, the microwave signals used to probe the cavity are extremely weak and need to be amplified before they can be processed by room temperature electronics. The challenge is that commercial amplifiers add significant noise which results in imperfect correlation between the measured output and the measurement backaction on the qubit. We use superconducting phase-sensitive parametric amplifiers to implement near noiseless amplification of a single quadrature of the microwave signal. 

\subsection{Parametric Amplification}
\label{sec:2:2}


There are several \idx{parametric amplifier} designs\cite{cast08, hatr11para, JPCNature}  but we will restrict our discussion to one based on a lumped element non-linear oscillator \cite{hatr11para}. The basic physics of parametric amplification in such a system can be understood by considering a driven, damped Duffing oscillator model \cite{vijayRSI} whose classical equation of motion is given by
\begin{eqnarray}
\frac{d^{2}\delta(t)}{dt^{2}}+ 2 \Gamma%
\frac{d\delta(t)}{dt}+\omega_{0}^2 \left(  \delta(t) - \frac{\delta(t)^3}{6}\right) \label{eqn:JBA}
=F\cos\left(  \omega_{d}t\right) \label{eq:nonlineq}.
\end{eqnarray}
Here $\omega_0$ is the linear resonant frequency for small oscillations, $\Gamma$ is the amplitude damping coefficient, $\delta$ is the gauge-invariant phase difference across the junction which is the dynamical variable of oscillator, and $\omega_d$ is the frequency of the harmonic driving term often referred to as the pump. 

\begin{figure}[t]
\centering
\includegraphics*[width=.9\textwidth]{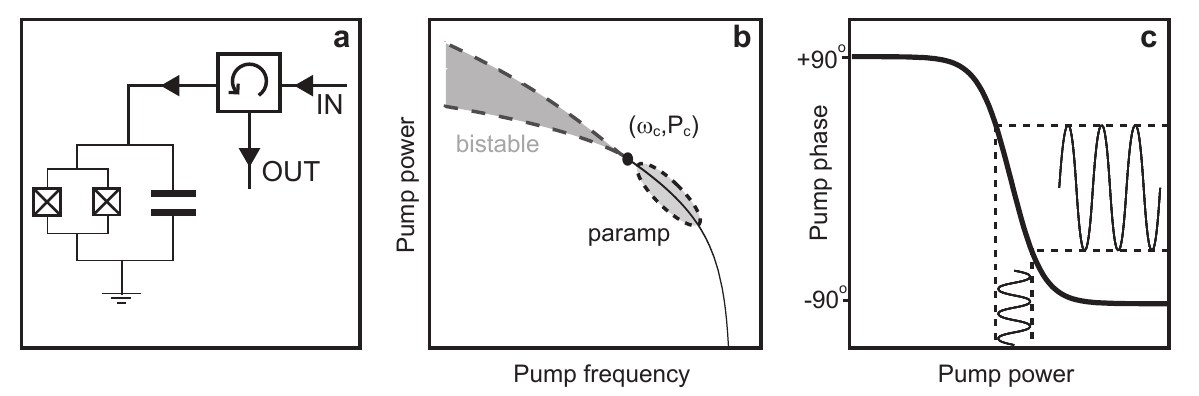}
%\picplace{5cm}{2cm} % Give the correct figure height and width in cm
\caption[]{(a) The basic circuit diagram of a Josephson junction based non-linear oscillator. A circulator is used to separate the amplified reflected signal from the incident signal. (b) Phase diagram of a driven non-linear oscillator. The solid line indicates the effective resonant frequency as a function of drive power. The dashed lines enclose the bistable regime of the non-linear oscillator. The oval marks the paramp biasing regime. (c) The reflected pump signal phase is plotted as a function of pump power and defines the transfer function of the paramp.}
\label{fig:paramp}       % Give a unique label
\end{figure}

Figure \ref{fig:paramp}(a) shows the basic circuit diagram of a Josephson junction based non-linear oscillator. The phase diagram of such a non-linear oscillator as a function of drive frequency and drive power is shown in Figure \ref{fig:paramp}(b). A characteristic feature of a \idx{non-linear oscillator} is that its effective resonant frequency is no longer independent of the amplitude of its oscillations. This is shown schematically by the black solid line in figure \ref{fig:paramp}(b) where the effective resonant frequency decreases with increasing driving power for Josephson junction based nonlinear oscillators. Beyond a critical drive frequency and power ($\omega_c$ and $P_c$), the driven response becomes bistable and we are not interested in that regime. The relevant regime for \idx{parametric amplification} is marked in Figure \ref{fig:paramp}(b) and is just before the onset of bistability. Because of the power dependence of the effective resonant frequency, it is possible to cross the resonance (black line) both in frequency and amplitude. Figure \ref{fig:paramp}(c) shows the phase of the reflected pump signal from the oscillator as a function of pump amplitude. This is essentially the transfer function of the amplifier and the pump power is chosen to bias the system in the steep part of this curve.

The signal to be amplified has the same frequency $\omega_d$  with an amplitude which is typically less than 1\% of the pump amplitude, is combined with the pump signal and sent to the nonlinear resonator.  Since the signal and pump are at the same frequency, one can define a phase difference between them. If the signal is in phase with the pump, then the combined amplitude changes significantly with the input signal which moves the bias point and consequently the reflected pump phase. This is essentially the mechanism of amplification, i.e. a small signal brings about a large phase shift in the large pump signal. However, if the signal is 90 degrees out of phase with the pump, then to first order, it has no effect on the bias point and consequently no change in the reflected pump phase. So the signals which are in phase with the pump get amplified while the quadrature phase signals get de-amplified. This allows one to selectively amplify the `Q' quadrature of the microwaves scattered from the cavity in a cQED measurement. In principle, such a phase-sensitive amplifier can amplify one quadrature without adding any additional noise to the signal \cite{hatr11para}, resulting in a signal-to-noise-ratio that is maintained after amplification and an output noise level that is entirely set by amplified zero point fluctuations associated with the coherent state. In practice, due to signal losses between the cavity and the amplifier and due to inefficiencies in the amplifier itself, one typically obtains an efficiency $\eta \sim 0.5$. In other words, the output noise is only about double the unavoidable quantum noise \cite{vija12,murc13traj}.

\subsection{Weak Measurement and Backaction}
\label{sec:2:3}

\begin{figure}[t]
\centering
\includegraphics*[width=.8\textwidth]{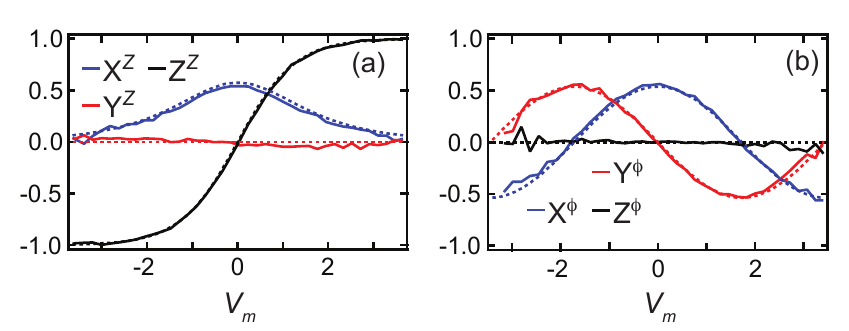}
%\picplace{5cm}{2cm} % Give the correct figure height and width in cm
\caption[]{Backaction of a weak measurement. (a)  When the `Q' quadrature of the scattered microwave signal is amplified a $\sigma_z$ measurement of the qubit occurs.   The dashed lines indicate the theoretical prediction from (\ref{eq:zback}) and the solid lines are the tomographic results conditioned on the measurement result $V_m$.  (b) A `$\phi$' measurement of the qubit is obtained by amplifying the `I' quadrature.  The dashed lines indicate the predictions from (\ref{eq:phiback}) and the solid lines are the conditioned tomography.  The data correspond to $\eta = 0.49$ and $S = 3.15$ and correspond to the experimental setup in \cite{murc13traj}.} 
\label{fig:weakmeas}       % Give a unique label
\end{figure}

A single \idx{weak measurement} $V_m$ is obtained by integrating the homodyne signal $V_Q(t)$ corresponding to the `Q' quadrature for a measurement time $\tau$, 
\begin{eqnarray}
V_m(\tau)  =\frac{2}{\tau \Delta V} \int_0^\tau V_Q(t)dt 
\label{eq:Vm}
\end{eqnarray}
where $\Delta V$ is the difference in the mean value of $V_Q(t)$ corresponding to the two qubit states. This implies that the mean value of $V_m$ is $\pm 1$ for the two qubit states. Measuring the `Q' quadrature of the scattered microwave signal in cQED using a phase-sensitive amplifier corresponds to a $\sigma_z$ measurement of the qubit. The corresponding \idx{backaction} pushes the qubit state towards one of the eigenstates of $\sigma_z$. This is shown as the solid black line in figure \ref{fig:weakmeas}(a) where $Z^Z = \langle\sigma_z\rangle |_{V_m}$ is plotted as a function of weak measurement result $V_m$ for an initial qubit state $(|0\rangle + |1\rangle)/\sqrt{2}$. Similary, the expectation values of $\sigma_x$ and $\sigma_y$ are also shown as solid blue and red lines respectively. The dashed line corresponds to the theoretical prediction given by \cite{koro11,murc13traj},
\begin{eqnarray}
Z^Z = \tanh\left(\frac{V_m S}{4} \right) ,  X^Z = \sqrt{1-(Z^Z)^2}e^{-\gamma \tau} ,  Y^Z = 0 \label{eq:zback}
\end{eqnarray}
where $\tau$ is the measurement duration, $S=64\tau\chi^2\nbar \eta / \kappa$ is the dimensionless measurement strength and $\gamma$ is the environmental dephasing rate of the qubit. To connect with the nomenclature used in section \ref{sec:1:2}, we note that $V_m$ is equal to the dimensional measurement result $V$ defined earlier whereas $S=4/a^2$ where $a^2$ is the variance of Gaussian measurement distributions.  If $\eta=1$, the state remains pure during the entire measurement process. Here, the data corresponds to $\eta=0.49$ and $S=3.15$.

It is important to note that even though the other quadrature `I' does not contain any information about $\sigma_z$, it does contain information about the photon number fluctuations in the coherent state \cite{koro11}. This implies that a measurement of the `I' quadrature will result in the qubit state rotating about the $Z$ axis in the Bloch sphere \cite{murc13traj} and will not push the state towards one of the eigenstates of $\sigma_z$. This can be seen in figure \ref{fig:weakmeas}(b) where the expectation values of $\sigma_{x,y,z}$ are plotted (solid lines) after a weak measurement with the paramp amplifying the `I' quadrature. Here $V_m$ is obtained by integrating the homodyne signal $V_I(t)$ corresponding to the `I' quadrature for a measurement time $\tau$. Note that this is not the same as measuring the $\sigma_x$ or the $\sigma_y$ operator and we label it as a `$\phi$' measurement since it results in rotation about the $Z$ axis in Bloch sphere. Consequently, there is no asymptotic value for $X^\phi$ or $Y^\phi$ and they evolve sinusoidally as, 

\begin{eqnarray}
X^\phi = \cos\left(\frac{V_m S}{4} \right)e^{-\gamma \tau} , \quad Y^\phi = -\sin\left(\frac{V_m S}{4} \right)e^{-\gamma \tau} ,\quad Z^\phi = 0 \label{eq:phiback}.
\end{eqnarray}
and $Z^\phi$ doesn't evolve and remains zero. 

A measurement using a \idx{phase-sensitive amplifier} truly enables one to measure one quadrature only while erasing the information in the other quadrature so that no observer can get access to that information. In fact, that information no longer exists and is not the same as ignoring the information in that quadrature which would lead to decoherence. So a single quadrature measurement of the `Q' quadrature only provides information about $\sigma_z$ and there are no photon number fluctuations. Similarly, when measuring the `I' quadrature, no information about $\sigma_z$ is available and hence $\langle\sigma_z\rangle$ does not change \cite{murc13traj}.

It is also possible to measure both quadratures simultaneously by using a phase preserving amplifier.  The parametric amplifier described above can be used in phase preserving mode by detuning the signal frequency from the pump frequency \cite{hatr11para} or one could use a two mode amplifier like the Josephson Parametric Converter (JPC) \cite{JPCNature}. In this case, each measurement gives two outputs corresponding to the `I' and `Q' quadrature \cite{hatr13}. This implies that one learns about the qubit state as well as the photon number fluctuations and consequently the measurement backaction results in the Bloch vector rotating around the $Z$ axis while it approaches one of the eigenstates of $\sigma_z$. Somewhat surprisingly, even in this case the qubit state can remain pure \cite{hatr13} throughout the measurement process provided the measurement efficiency $\eta = 1$.

\section{Quantum Trajectories}
\label{sec:3}
\idx{Quantum trajectories} were first introduced as a theoretical tool to study open quantum systems \cite{carm93,gardinerbook,dali92,gard92,scha95}. Rather than describe an open quantum system by a density matrix, which for a $N$-dimensional Hilbert space requires $N^2$ real numbers and requires solving the master equation, the evolution of a pure state (which requires only $N$ complex numbers), can be repeatedly calculated to determine the evolution of $\rho(t)$.  The quantum trajectory formalism thus assumes that the evolution of an open (and therefore mixed) quantum system can be expressed as the evolution of several, individual, pure quantum trajectories. 

\subsection{Continuous Quantum Measurement}

\begin{figure}[t]
\centering
\includegraphics*[width=.7\textwidth]{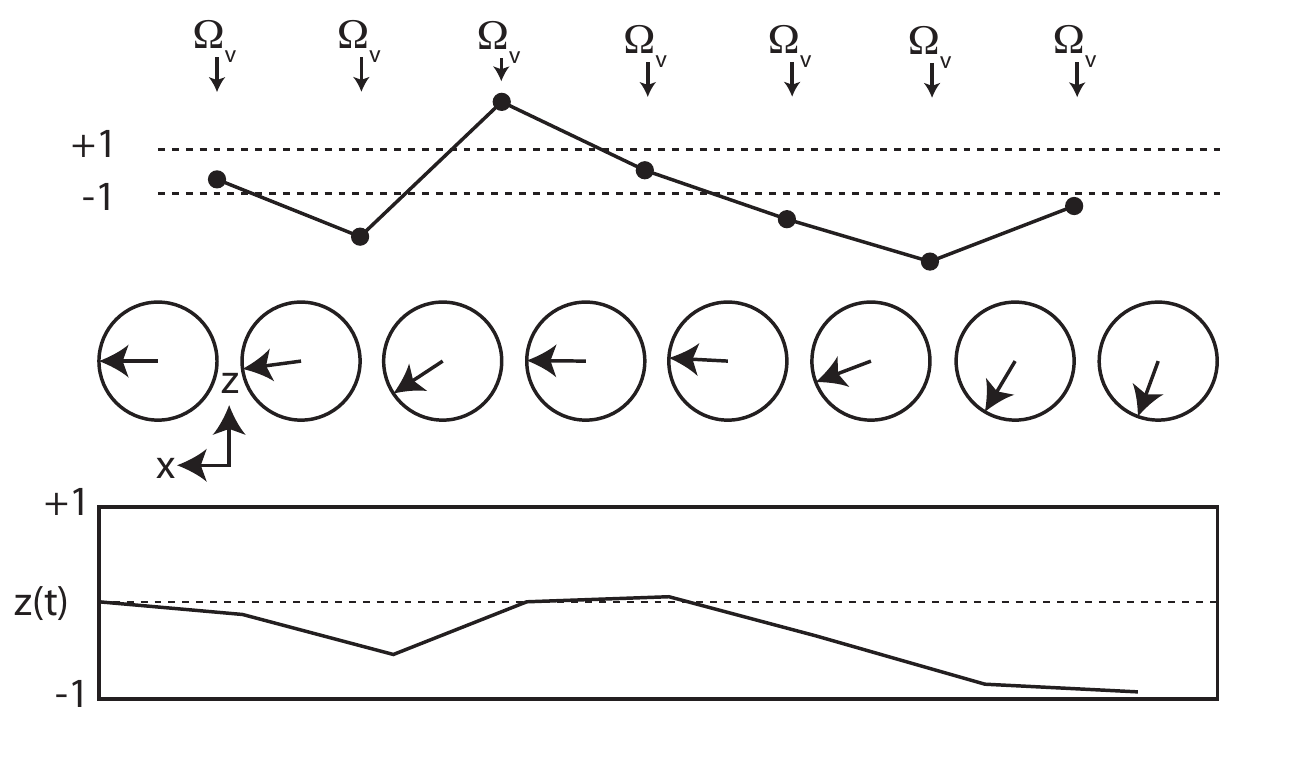}
%\picplace{5cm}{2cm} % Give the correct figure height and width in cm
\caption[]{A sequence of weak measurements leading to a quantum trajectory of the qubit state on the Bloch sphere. The qubit is initially along $x = 1$ and each measurement imparts conditional dynamics on the state.}
\label{fig:seq}       % Give a unique label
\end{figure}

	The process of \idx{continuous quantum measurement} can be built out of a sequence of discrete weak measurements as sketched in figure \ref{fig:seq}.  As we have already established, each of these measurements induces a specific conditional backaction on the quantum state. If several of these weak measurements are conducted in series then the discrete time evolution of the quantum state can be determined.  The cQED measurement apparatus that we consider forms a continuous probe of the quantum system, however since the cavity has a finite bandwidth $\kappa$, the measurement signal is correlated at times less than $\sim 1/\kappa$ and it makes sense to bin the measurement signal in time steps $\Delta t\sim1/\kappa$.  For a cavity with $\kappa/2\pi = 10$ MHz, this correlation time is roughly 16 ns, so the continuous measurement record is discretized in similar time steps.  These discrete, weak measurements can be considered to be continuous in the limit where the discretization steps are much smaller than the characteristic measurement time $\tau_c = \kappa/16 \pi \chi^2 \nbar \eta$ which was defined in section \ref{sec:1:2}. Typically, experiments operate with $\tau_c \simeq 1 \ \mu$s, such that $\Delta t\ll \tau_c$.
	
\begin{figure}[t]
\centering
\includegraphics*[width=.9\textwidth]{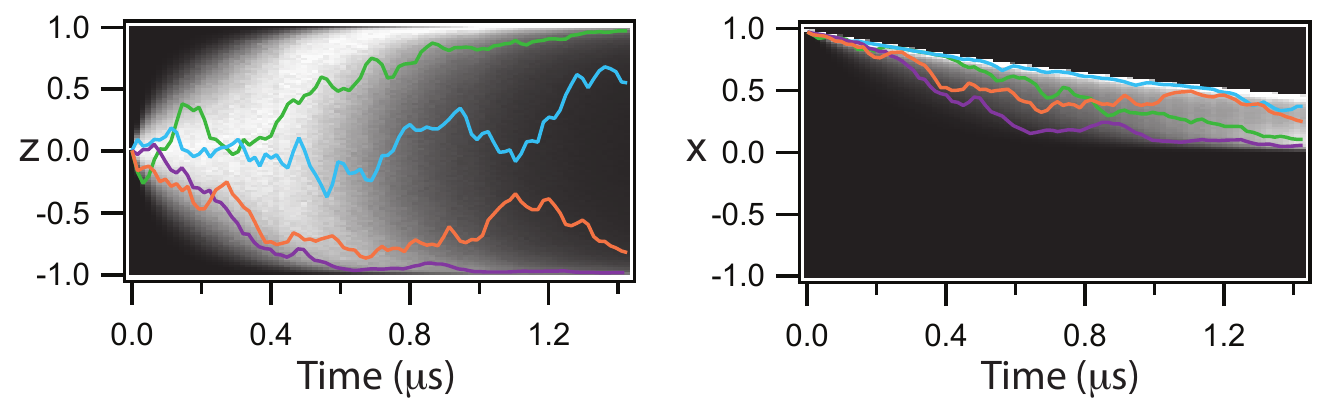}
%\picplace{5cm}{2cm} % Give the correct figure height and width in cm
\caption[]{Individual quantum trajectories of the qubit state as given by $z =\langle \sigma_z \rangle$  and $x =\langle \sigma_x \rangle$ from an initial state $+x$.  Four different trajectories are shown in color on top of a greyscale histogram indicates the relative occurrence of different states at different times. }
\label{fig:ztraj}       % Give a unique label
\end{figure}

	A single experimental iteration results in a continuous measurement signal $V(t)$ which is subsequently binned into a string of measurement results $(V_i, V_{i+1}, V_{i+2}$). Given the initial state of the qubit, the state is updated at each time step $t_i$, either based on the \idx{stochastic master equation} \cite{jaco06, tan15}, or a Bayesian argument \cite{murc13traj}.  This leads to a discrete time trajectory $x(t),y(t),z(t)$.  Figure \ref{fig:ztraj} displays several of these trajectories for the qubit initialized in the state $x=+1$.  Because each experiment results in a different measurement signal $V(t)$, each trajectory is different.
	
\begin{figure}[t]
\centering
\includegraphics*[width=.7\textwidth]{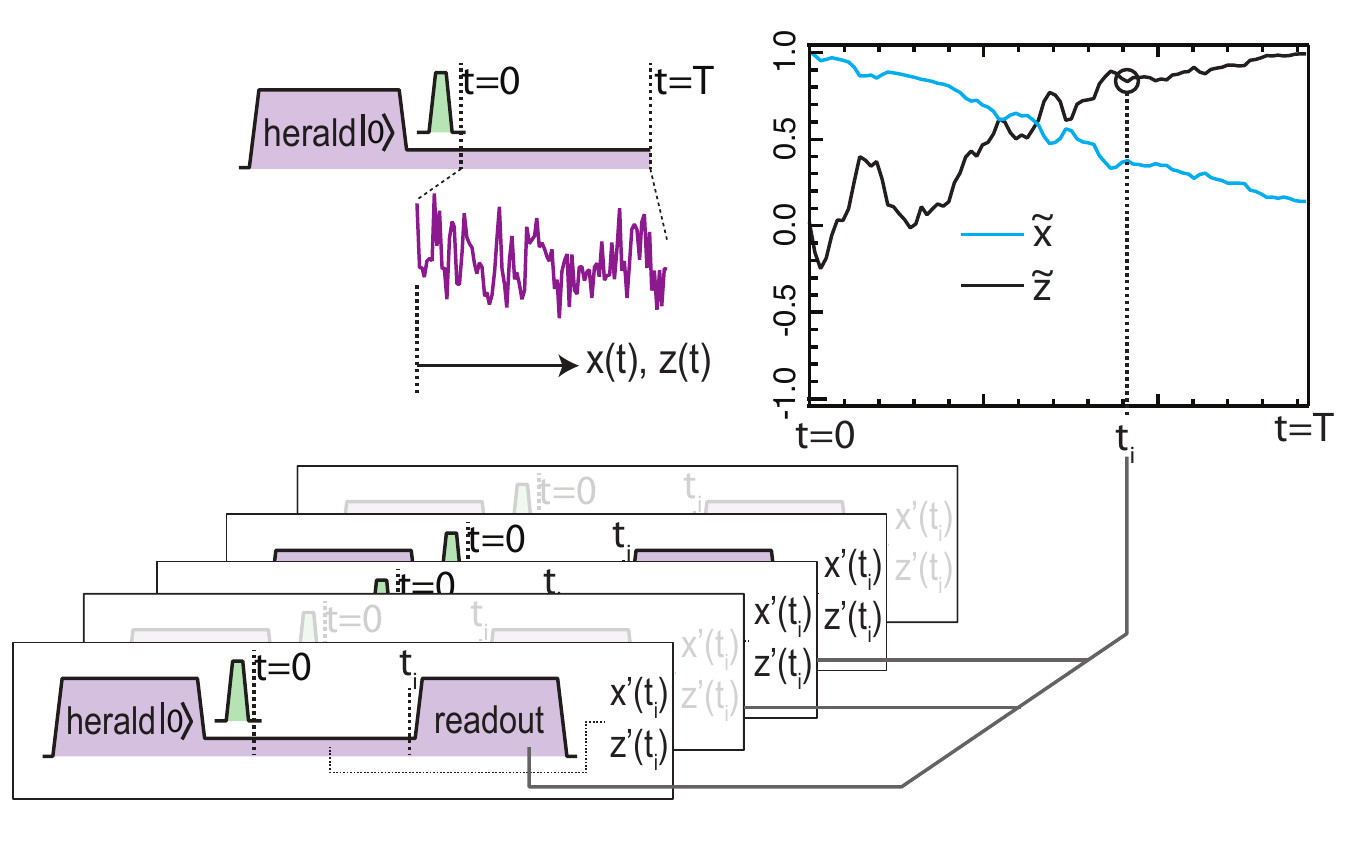}
%\picplace{5cm}{2cm} % Give the correct figure height and width in cm
\caption[]{Tomographic reconstruction of the trajectory. To reconstruct the target trajectory at time $\tilde{z}(t_i)$ several experiments are performed with a projective measurement of $\sigma_z$ at time $t_i$. For each of these, the state $(x'(t_i), z'(t))$ is calculated and if it matches the target trajectory the outcome of the projective measurement is included in the average.}
\label{fig:1}       % Give a unique label
\end{figure}

	To verify that these trajectories, which are conditional on a single measurement signal, are correct we have to prove that at every point along the trajectory the conditional state makes correct predictions for the outcome of any measurements that can be performed on the system.  To accomplish this, we perform conditional \idx{quantum state tomography} at discrete times along the trajectory.  We denote a single trajectory $\tilde{x}(t)$, $\tilde{z}(t)$ as a ``target'' trajectory.  This trajectory makes predictions for the mean values of measurements of $\sigma_x$ and $\sigma_z$ versus time. To create a tomographic validation of these predictions at a specific time $t_i$ we perform several experiments with a weak measurement duration of $t_i$ that conclude with one of three tomographic measurements. For each of these experiments, we calculate the trajectory $x'(t_i),z'(t_i)$ and if $x'(t_i) = \tilde{x}(t_i)\pm \epsilon$ and $z'(t_i) = \tilde{z}(t_i)\pm \epsilon$ the outcomes of the projective tomographic measurements are included in the average trajectory $x(t), z(t)$.  We find that the target trajectory and tomographic trajectory are in close agreement.  
	
	If we did not condition our trajectories on the measurement signal, for example if $\eta = 0$, then the trajectories would simply follow the ensemble evolution as described by a standard Lindblad master equation.  In this case, each trajectory would be the same, yet would still make correct predictions for the outcome of projective measurements performed on the system.  While this unconditioned evolution makes correct predictions, it also quickly takes an initial pure state to a mixed state.  However, if $\eta \sim 1$, the conditional quantum state retains substantial purity for all time and makes correct predictions for measurements on the system.
	
\subsection{Unitary Evolution}
	
	So far in this section we have considered the case of quantum non-demolition (QND) measurement, in that the weak measurements we perform $\propto \sigma_z$ commute with the qubit evolution Hamiltonian. Since the measurements are QND, all the measurements commute and only the integrated measurement signal is necessary for the conditional state evolution.  In this regime, the measurements can be treated simply in terms of classical probabilities, since the evolution only depends on the state populations, and the measurement signal $\propto \sigma_z$ also depends only on the populations.

\begin{figure}[t]
\centering
\includegraphics*[width=.9\textwidth]{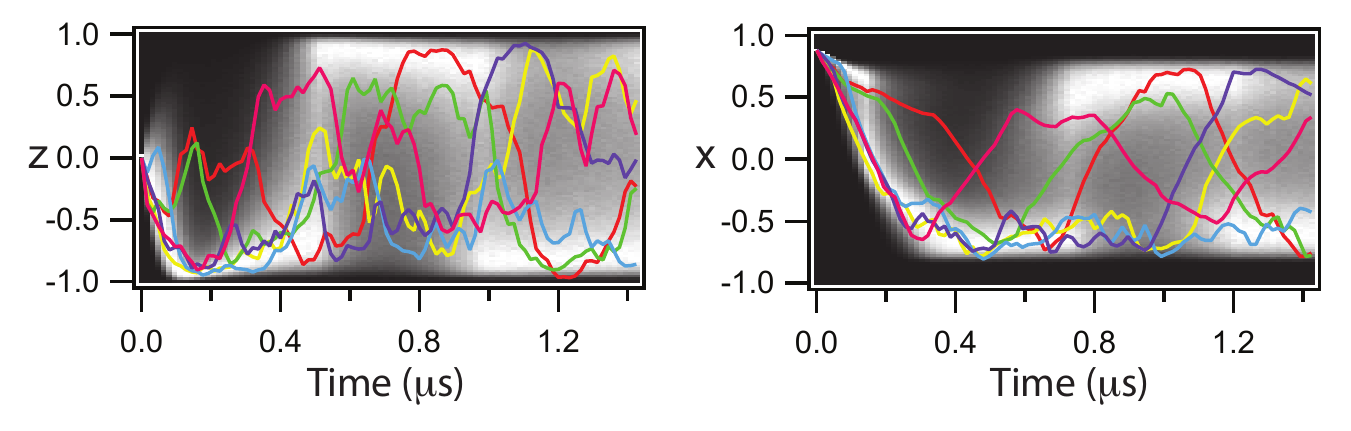}
%\picplace{5cm}{2cm} % Give the correct figure height and width in cm
\caption[]{Quantum trajectories in the presence of unitary driven evolution.  Individual quantum trajectories of the qubit state as given by $z =\langle \sigma_z \rangle$ from an initial state $+x$.  Six different trajectories are shown in color on top of a greyscale histogram indicates the relative occurrence of different states at different times. }
\label{fig:comp}       % Give a unique label
\end{figure}

	The situation becomes much more rich if we allow for unitary evolution of the qubit state that does not commute with the measurement operators.  To accomplish this, we drive the qubit resonantly to induce Rabi oscillations. This drive is described by the Hamiltonian, $H_R = \hbar \Omega \sigma_y/2$. In this case, the unitary evolution is included in the stochastic master equation, 
	\begin{align}
\frac{d \rho}{dt} = -\frac{\mathrm{{\bf i}}}{\hbar}[H_\mathrm{R},\rho] + k(\sigma_z \rho \sigma_z - \rho)
+ 2 \eta k( \sigma_z \rho +\rho \sigma_z - 2 \mathrm{Tr}(\sigma_z \rho) \rho) V_t. \label{eq:rho}
\end{align}
The Rabi drive turns coherences into populations and vice versa, causing the measurement signal to depend on the coherences of the qubit. Such evolution is fully quantum and reveals interesting features regarding the competition between unitary dynamics and measurement dynamics.   Figure \ref{fig:comp} displays several trajectories which exhibit this competition between measurement and driven evolution. The trajectories are oscillatory, but distorted by the stochastic backaction of the measurement, approaching jump-like behavior expected for the regime of quantum jumps. This figure also highlights how state tracking maintains the purity of the state in comparison to the full ensemble evolution.
	
\subsection{The Statistics of Quantum Trajectories}

We have so far demonstrated the ability to track individual quantum trajectories which evolve in response to measurement dynamics (wave function collapse) and also in competition with unitary driven evolution.  These trajectories are stochastic in the sense that the evolution of each trajectory is different.  But, what statements can we make about these trajectories in general?  Clearly, the evolution associated with weak measurement would be in some way different than the dynamics of quantum jumps \cite{vija12}, but how do we quantify and characterize trajectories?  

In order to characterize the general properties of trajectories we need a method to look at ensembles of trajectories but that does so in a way that depends on the individual trajectories.  To accomplish this, we consider the sub-ensemble of trajectories that start and end at certain points in quantum phase space.  This sub-ensemble is pre- and post-selected in that all the trajectories start from the same initial state (pre-selection) and we then post-select a sub-ensemble that ends in a specific final state. Given these pre- and \idx{post-selected} trajectories we can now examine aspects of the sub-ensemble. One such property is the most-likely path that connects the initial and final states.  This particular choice is of interest because such most-likely paths can be calculated with a stochastic action principle for continuous quantum measurement which maximizes the total path probability connecting quantum states \cite{chan13}.  Experiments \cite{webe14} show good agreement with the theoretical \idx{most-likely path} and the predicted path from theory, thus validating the theory which may be applicable in other quantum control problems. This analysis gives insight into the dynamics associated with open quantum systems, with applications in quantum control and state and parameter estimation.

\subsection{Time-symmetric State Estimation}

The examination of pre- and post-selected quantum trajectories raises the notion of \idx{time symmetry} in quantum evolution and quantum measurement  \cite{wana55,ahar64,ahar10, ahar11}.  We have so far used the quantum state as predictive tool, that is, at a time $t$ the quantum state described by $\rho(t)$ makes correct predictions for the probabilities of the outcomes of measurements performed at time $t$ and the associated mean values of these observables.  However, after a measurement is performed, the quantum state may continue to evolve due to further probing and unitary driving, and we may ask at some later time $T>t$ what is the probability for the outcome of that measurement \emph{in the past} given the results of later probing.  

Consider a measurement scenario where two experimenters monitor the evolution of a qubit and track its quantum state $\rho(t)$. At time $t$ one experimenter makes a measurement of the qubit but locks the result ``in a safe".  The second experimenter then continues to monitor the qubit and at a later time $T$ the second experimenter wants to guess the outcome of the measurement whose result is locked in the safe.  Clearly more information is available if the second experimenter accounts for information about the qubit obtained after the first measurement, and if this experimenter can correctly account for those results, he will be able to make more confident predictions for the result in the safe.  Stated simply, the second experimenter must determine what result is most likely to be locked in the safe given the subsequent measurement signal.

One can show \cite{gamm13} that at time $T$, the second experimenter's hindsight prediction for the measurement performed in the past is given by, 
\begin{eqnarray}
P_p(m)  = \frac{\mathrm{Tr}(\Omega_m \rho(t) \Omega_m^\dagger E(t))}{\sum_m \mathrm{Tr}(\Omega_m \rho(t) \Omega_m^\dagger E(t))}, \label{eq:pqs}
\end{eqnarray}
which describes the probability of obtaining outcome $m$ from the POVM measurement $\Omega_m$  performed by the first experimenter.  Here $\rho(t)$ is the usual quantum state propagated forward in time until time $t$ and $E(t)$ is a similar matrix which is propagated backwards from time $T$ to time $t$ using a similar method for the calculation of $\rho(t)$.  The matrix $E(t)$ has recently been calculated for experimental data and demonstrated that Eq.(\ref{eq:pqs}) makes correct and indeed more confident predictions for the outcome of measurements performed in the past \cite{ryba14,tan15}.

\section{Analog Feedback Stabilization: Rabi Oscillations}
\label{sec:4}

In this section, we will explore how one can use the continuous measurement record obtained using weak measurements to modify the behavior of the quantum system being monitored with the help of \idx{feedback}. Unlike feedback in classical systems, one has to worry about the random backaction associated with the measurement of a quantum system. However, as explained in previous sections, the state of a quantum system can be monitored perfectly using weak measurements if the initial state is known and the measurement efficiency is unity. Even though the evolution of the quantum system is random and unpredictable, it is still knowable by monitoring the weak measurement output. This can also be done in the presence of any additional unitary evolution as explained in \ref{sec:3}.2. However, decoherence processes which are always present will tend to take the system away from a desired state one might want. We will now describe how one can use quantum feedback to prevent the quantum system from deviating from a desired state. We will consider the particular case of feedback control in a resonantly driven qubit undergoing Rabi oscillations \cite{korotkov_dir_fb}.

\subsection{Weak Monitoring of Rabi Oscillations}

Let us first look at Rabi oscillations more carefully. The state of a resonantly driven qubit evolves sinusoidally between its two states with a rate $\Omega_R$ which depends on the strength of the resonant drive. To be specific, for a qubit state $\alpha(t) |0\rangle + \beta(t) |1\rangle$, $|\alpha(t)|^2 = \sin^2(\Omega_R t/2)$. These oscillations in the qubit state probability are called Rabi oscillations. In the absence of any decoherence, given the initial state and the Rabi frequency $\Omega_R$, we can predict the qubit state at any future time. One can equivalently say that the phase of the \idx{Rabi oscillations} is known and remains unchanged with time. However, decoherence processes will introduce errors in this deterministic evolution and over some characteristic time scale, the phase of the Rabi oscillations will diffuse. 

\begin{figure}[b]
\centering
\includegraphics*[width=.6\textwidth]{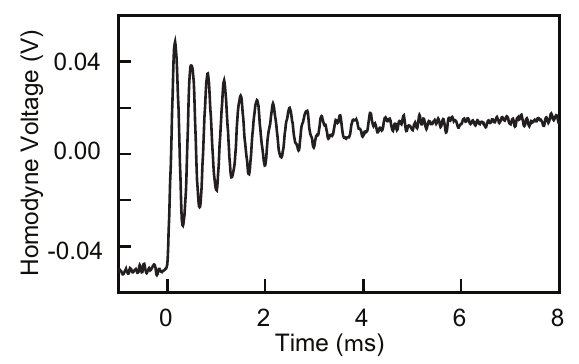}
%\picplace{5cm}{2cm} % Give the correct figure height and width in cm
\caption[]{Rabi oscillations obtained using ensemble averaged weak measurements. The decay constant is set by both environmental decoherence and measurement induced decoherence.}
\label{fig:weakrabi}       % Give a unique label
\end{figure}

In a typical Rabi oscillation experiment, the qubit is initialized in the ground state and resonantly driven for a fixed duration of time ($\tau_R$) followed by a projective measurement. This process is repeated many times to obtain the ensemble averaged qubit state $\langle\sigma_z\rangle$. A plot of $\langle\sigma_z\rangle$ as a function of $\tau_R$ yields decaying sinusoidal oscillations where the decay constant depends on the decoherence in the system. The decaying oscillations are an indication of the diffusion of the Rabi oscillation phase with time. Here, qubit driving and projective measurement are never done simultaneously. However, one can drive the qubit while measuring it weakly and still obtain ensemble averaged Rabi oscillations. As discussed in \ref{sec:1:2}, since the weak measurement output can be thought of as a noisy estimate of $\langle\sigma_z\rangle$, an ensemble average of the weak measurement signal in the presence of Rabi drive also yields decaying oscillations. Figure \ref{fig:weakrabi} shows Rabi oscillations obtained using weak measurements. However, one important difference is that the decay constant now depends on both environmental decoherence and measurement strength. This additional measurement induced decoherence is a consequence of the ensemble averaging where we ignore the individual results of the weak measurements. This is in contrast to the oscillatory quantum trajectories shown in Figure \ref{fig:comp} with simultaneous Rabi driving and weak measurement.

We will now discuss a \idx{feedback protocol} \cite{korotkov_dir_fb} which corrects for the phase diffusion of Rabi oscillations and prevents the decay of Rabi oscillations. In principle, one can do a full reconstruction of the quantum state \cite{haroche_fb} to estimate the feedback signals required. However, to do that in real-time is experimentally challenging. Instead, we use classical intuition in this feedback protocol motivated by  phase-locked loops (PLL) used to stabilize classical oscillators. In a PLL, one compares the phase of an oscillator with that of a reference signal. Any phase error is then corrected by creating a feedback signal proportional to the error which controls the oscillator frequency. Essentially, if the oscillator is lagging in phase, then the feedback signal increases the frequency and vice-versa. 

\begin{figure}[t]
\centering
\includegraphics*[width=.9\textwidth]{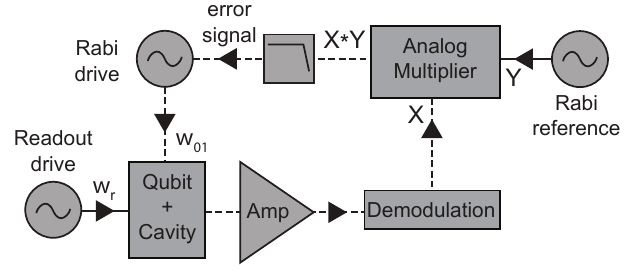}
%\picplace{5cm}{2cm} % Give the correct figure height and width in cm
\caption[]{Simplified feedback setup for stabilization of Rabi oscillations.}
\label{fig:fbsetup}       % Give a unique label
\end{figure}

\begin{figure}[t]
\centering
\includegraphics*[width=1.0\textwidth]{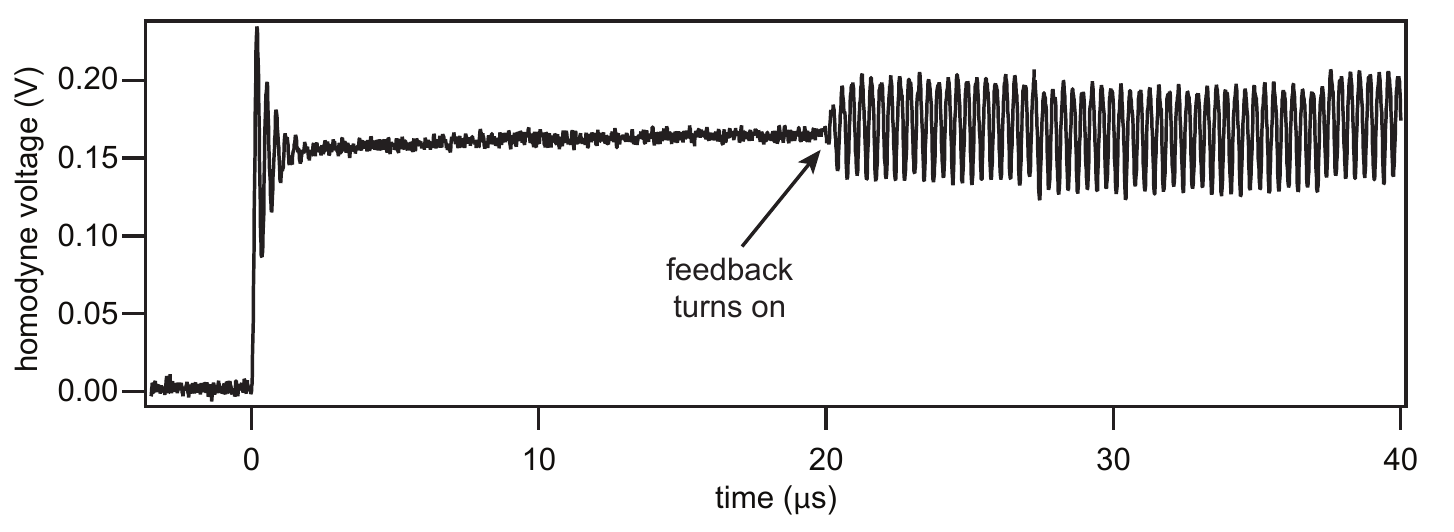}
%\picplace{5cm}{2cm} % Give the correct figure height and width in cm
\caption[]{Feedback stabilized Rabi oscillations. Initially the ensemble averaged Rabi oscillations decay, primarily due to measurement induced decoherence. Feedback is turned on later and results in the oscillations recovering as the phase of Rabi oscillations synchronize with the reference signal.}
\label{fig:rabifeedback}       % Give a unique label
\end{figure}

We can now apply the same idea to our weakly monitored qubit in the presence of Rabi driving \cite{vija12}. The basic feedback setup is shown in Figure \ref{fig:fbsetup}. A reference signal at the Rabi frequency is multiplied with the weak measurement signal and low pass filtered to create the error signal. Since the weak measurement signal is a noisy oscillatory signal corresponding to the oscillating qubit state, the error signal is proportional to the deviation in phase of the Rabi oscillations with respect to the reference signal. The error signal is used to  control the amplitude of the Rabi drive which in turn controls the Rabi frequency just as in a PLL. Figure \ref{fig:rabifeedback} shows the effect of such a feedback signal which is turned on after a time much greater than the decay constant of the Rabi oscillation in the absence of feedback. On can clearly see that the ensemble averaged oscillations recover when the feedback is turned on and stabilize to a fixed amplitude. As long as the feedback is on, the ensemble averaged oscillations will never decay. This implies that the Rabi oscillations have synchronized with the reference signal and the phase diffusion has been reduced due to feedback though not completely eliminated. Note that the slow drift in the mean level of the signal is due to finite probability of getting excited into the second excited state of the transmon qubit \cite{vija12}.

\begin{figure}[t]
\centering
\includegraphics*[width=1.0\textwidth]{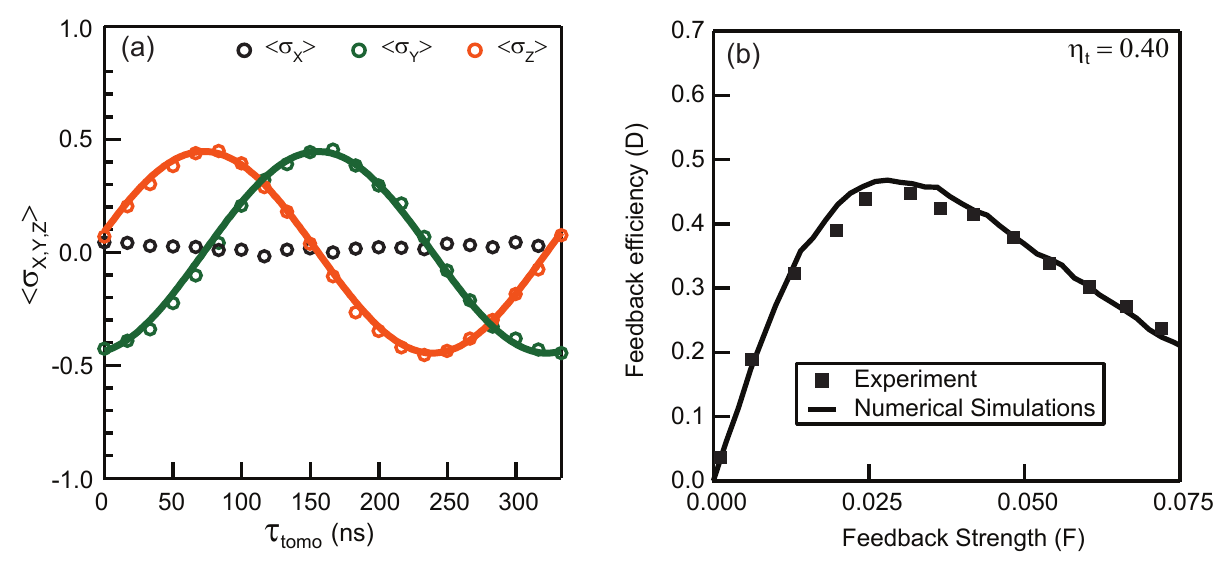}
%\picplace{5cm}{2cm} % Give the correct figure height and width in cm
\caption[]{(a) Quantum state tomography of feedback stabilized Rabi oscillations showing $\langle\sigma_{X}\rangle$,  $\langle\sigma_{Y}\rangle$,  and $\langle\sigma_{Z}\rangle$  for one full Rabi oscillation. (b) Feedback efficiency is plotted a function of feedback strength.}
\label{fig:tomo}       % Give a unique label
\end{figure}

To ensure that the stabilized oscillations are not an artifact of the measurement setup, \idx{quantum state tomography} was used to verify the quantum nature of the stabilized oscillations. Figure \ref{fig:tomo}(a) shows a plot of $\langle \sigma_{X}\rangle$, $\langle\sigma_{Y}\rangle$, and $\langle\sigma_{Z}\rangle$ for one full stabilized Rabi oscillation. The magnitude of these oscillations do not reach $\pm1$ indicating that the
synchronization is not perfect and the phase diffusion of the Rabi oscillations has not been completely eliminated. The data shown is the best synchronization we obtained in this experiment corresponding to a feedback efficiency $D=0.45$ which is approximately given by the amplitude of the oscillations in  $\langle\sigma_{Z}\rangle$ or  $\langle\sigma_{X}\rangle$. The best synchronization is obtained for an optimal feedback strength $F$ as is evident from Figure \ref{fig:tomo}(b) which shows a plot of $D$ vs $F$ (solid squares). The dimensionless feedback strength $F$ is essentially the feedback loop gain and is controlled by the amplitude of the reference signal.

There are two main factors in this experiment which result in $D<1$. The first one is the total \idx{measurement efficiency} $\eta_t$ which had a  value of 0.4 in this experiment. This efficiency has two contributions given by $\eta_t = \eta \ \eta_{\mathrm{env}}$. Here $\eta$ is the measurement efficiency due to the detector as introduced in \ref{sec:1:2} and is set by the excess noise introduced by the amplification chain. The other term $\eta_{\mathrm{env}} = (1+\Gamma_{\mathrm{env}} / \Gamma_m)^{-1}$ takes into account the environmental dephasing $\Gamma_{\mathrm{env}}$, where $\Gamma_m$ is the measurement induced dephasing. In order for feedback stabilization to work well, one needs to ensure that the environmental dephasing is small compared to the measurement induced dephasing i.e.  $\eta_{\mathrm{env}} \rightarrow 1$. In other words, you want the measurement induced disturbance to dominate over the environmental disturbance since the measurement output can then be used to correct for those. The second factor affecting the feedback efficiency is the loop delay. Since the qubits are operating inside a dilution refrigerator while the feedback electronics are operating at room temperature, there is a delay in the feedback signal which results in inefficient synchronization. The solid line in Figure \ref{fig:tomo}(b) is obtained using numerical simulations including the effect of feedback delay and shows good agreement with the experimental data. While one might be tempted to increase measurement strength arbitrarily to approach $\eta_t=1$,  feedback delay and finite feedback bandwidth leads to an optimal value of measurement strength for maximizing feedback efficiency. This is because stronger measurement requires faster feedback which is limited by the bandwidth of the feedback loop.

This experiment demonstrates the use of continuous measurement and feedback to stabilize Rabi oscillations in a qubit. The simple feedback protocol which is based on classical intuition works successfully because the feedback signal achieves near perfect cancellation of the random measurement backaction for an optimal value of $F$. This technology can provide another route for quantum error correction based on weak continuous measurements and feedback \cite{qerrcorr,parity_cQED} with a potential advantage in situations where strong measurements can cause qubit state mixing \cite{slichter_dd_prl}. Such techniques also offer the possibility of  measurement based quantum control for solid-state quantum information processing \cite{Hofmann_qcontrol,state_stable,Gillett-2010,Korotkov_entang,state_pur_jacobs,jacobs_adaptive_msmt,cook-2007}. 

\section{Conclusion}
\label{sec:5}

Weak measurements realize a flexible method of implementing active feedback in quantum systems. With superconducting circuits that operate at microwave frequencies, feedback fidelity is currently limited by the overall measurement efficiency of the amplification chain. In typical setups, the amplifier is housed in a separate cryo-package and inefficiencies result from losses in cable connectors and passive components such as circulators which add directionality to the signal path. One promising avenue to overcome this limitation is to use parametric devices that can be directly integrated on-chip with the qubit. Such types of circuits use a combination of complex pumping techniques at different frequencies \cite{ranzani2015} or multiple cavity modes to isolate the amplifier bias form the qubit circuit \cite{pari2015}. Additionally, as more sophisticated feedback routines are developed, particularly sequences which involve digital processing of the measurement data, then loop delays resulting from long data paths and latencies in classical electronics must also be taken into account. Wiring complexity and dead time maybe minimized by integrating quantum circuits with cryogenic classical logic, either superconducting or semiconducting.

Another frontier to be explored in superconducting circuits is the use of feedback and control in multi-qubit arrays. For example, in such an architecture, one can imagine simultaneously performing weak measurements on each qubit. The resulting joint-state information can potentially be used to reconstruct the initial state of the array from families of quantum trajectories, realizing another form of state tomography that capitalizes on the dense information embedded in the correlations of an analog weak measurement. Such techniques can also be extended for Hamiltonian parameter estimation, parity measurements, and measurement based error correction. In essence, simultaneous probing of a quantum many body system parallelizes both `read' and `write' operations. 

\printindex

 % \bibliographystyle{spphys} 
 %\bibliography{ch7_references-1}

\end{document}